Science

# Wind-Enhanced Interaction of Radiation and Dust (WEIRD) and the Growth and Maintenance of Local Dust Storms on Mars

Scot C. Randell Rafkin[1], Aaron Rothchild[1,2], Timothy I. Michaels[1],and Roger A. Pielke, Sr.[2]

[1]Dept. of Space Studies, Southwest Research Institute, Boulder, CO, 80302, USA, rafkin@boulder.swri.edu; [2]Cooperative Institute for Research in Environmental Sciences, University of Colorado, Boulder, CO, 80309, USA

### Abstract

**Background:** A radiative-dynamic positive feedback mechanism (Wind Enhanced Interaction of Radiation and Dust: WEIRD) for localized Mars dust disturbances was previously found to operate in highly idealized numerical experiments. The ability for this mechanism to operate under more realistic conditions is investigated.

**Method:** Numerical simulations of the proposed Mars Science Laboratory landing site at Mawrth Valles are used to test for the presence and quantitative effect of the radiative-dynamic WEIRD feedback mechanism under more realistic conditions. Using observed surface properties (e.g., albedo, thermal inertia, topography), general circulation model data for initialization and boundary conditions, a fully active dust cycle with a mass bin model representation, and a radiative transfer scheme that interacts fully with the aerosol model, mesoscale simulations are performed and analyzed to quantify the impact of local and regional atmospheric dust loading on the local and regional scale circulations. Comparisons between cases where lifted dust is radiatively active and radiatively passive elucidate the importance of the dust radiative forcing on the thermodynamic and kinematic structure of the atmosphere.

**Conclusion:** The WEIRD feedback mechanism does operate under realistic conditions, although it can be masked and diminished by a variety of other forcing mechanisms. Globally increased dust loading is found to accelerate the local winds while simultaneously diminishing the impact of local physiographical forcing. Local enhancements of dust produce a thermal and dynamical response that resembles many of the essential features seen in the idealized experiments. The development of a warm core low, rotational wind tendencies and convergence boundaries intersecting at the center of the strongest dust storms are consistent with WEIRD. Local and regional storms are effective at producing elevated dust layers above the boundary layer aided by the radiative forcing of the dust. Ubiquitous and persistent thermal circulations associated with topography can also inject dust into the free atmosphere above the planetary boundary layer, but they are less efficient than the dust storms. High concentrations of dust in the lowest levels of the atmosphere produce a significant and dramatic heating of the ground and the near-surface air despite greatly reduced insolation at the surface.

## Introduction

In a previous study, a positive radiative-dynamic feedback process was shown to operate under specific conditions within idealized local dust disturbances on Mars (Rafkin 2009). Here, we extend the study to more realistic conditions by analyzing the local response of the atmosphere to global, regional, and local dust storms using a radiatively active dust cycle within a mesoscale model (Rafkin et al 2001).

The question as to whether this feedback mechanism can operate under realistic conditions, as opposed to the idealized conditions imposed in the earlier study, remains open. The following questions are also unanswered: How does dust modulate the local and regional scale circulations?  What are the conditions inside a dust storm near regions of active lifting?  Can the evidence of these circulations be found in orbital imagery?

## The Radiative-Dynamic Feedback Mechanism

Briefly, the positive feedback mechanism works as follows (Fig. 1):

1) Wind lifts dust from the surface into the atmosphere;

2) The increased atmospheric dust load results in increased radiative heating of the atmosphere during the day, or less radiative cooling during the night, thereby producing a relatively warm region on the scale of the lifted dust;

3) Surface pressure is hydrostatically lowered in the warm region, which leads to an amplification of the low-level horizontal pressure gradient force;

4) The increased horizontal pressure gradient results in stronger winds, which lift more dust, and thus completes the positive feedback loop.

In Rafkin (2009) the feedback process was shown to operate in a manner similar to the Wind Induced Sensible Heat Exchange (WISHE) process—essentially a Carnot Engine—that explains the intensification of tropical cyclones on Earth (Emanuel 1991). For the case of Mars dust disturbances, it is a Wind Enhanced Interaction of Radiation and Dust (WEIRD), as it is the radiative heating of lifted dust that drives the feedback on Mars. With tropical cyclones, it is the latent heat flux and the conversion to sensible heat in deep convective clouds that drives the Carnot-like engine. In both cases, however, the storm environment and surface thermodynamic properties strongly influence the intensification process. Barotropic environments and steep lapse rates are favorable while strongly sheared environments and environments with cold air near the surface or warm air aloft are not. The availability and amount of dust available for lifting is important for dust disturbances just as the availability of warm water is important for hurricanes. Not mentioned in Rafkin (2009), is the albedo of the surface which could play a role, because the surface absorption and reflection of solar irradiance (including the change of albedo over time from dust deposition and removal) influence the radiative forcing of the aerosols above the surface.

Beyond environmental factors, latitude plays an important role in the feedback process. Latitude exerts dynamic control via the coriolis torque and also determines the available solar input as a function of aerocentric longitude. Low latitudes have the greatest solar input, but lack the planetary vorticity to develop and sustain intense, balanced circulations. High latitudes are dynamically favorable, but lack the solar forcing. The subtropics provide the most favorable combination of solar forcing and coriolis forcing.

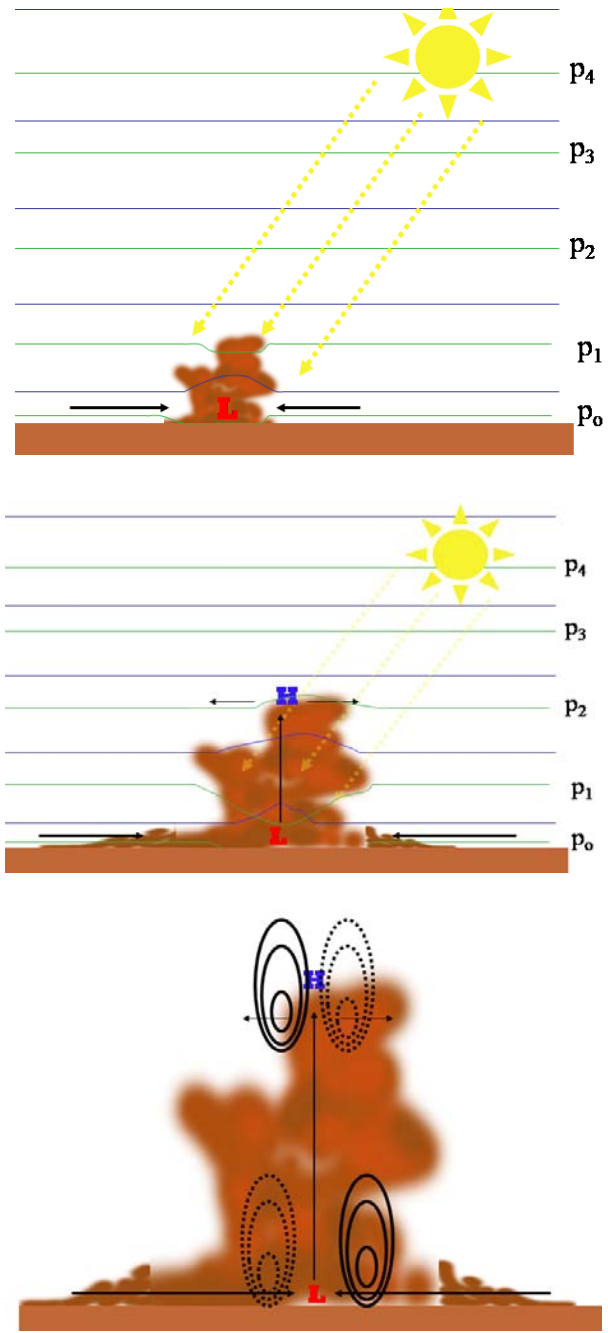

**Figure 1.** Schematic illustration of the wind-enhanced interaction of radiation and dust (WEIRD) radiative-dynamic positive feedback mechanism.  The heating of an initial dust disturbance (top) induces a hydrostatic lowering of pressure and an inward acceleration of winds. The increased winds lift more dust and upper level divergence develops from mass continuity (middle). Over time, a balanced dynamical response, consistent with a classic warm core low pressure system is established (bottom).  A strong cyclonic circulation is produced near the surface and an anticyclone is present aloft (fig1.jpg).

## Model Description

The model used to explore WEIRD in a more realistic setting is the Mars Regional Atmospheric Modeling System (MRAMS), the core of which is described by Rafkin et al. (2001). Two important improvements since Rafkin (2009) are applied for this study. First, the dust lifting parameterization of Michaels (2006) is used. Second, dust is treated via a dual tracer field that permits both a background distribution and a foreground perturbation distribution upon the background dust field.

### Dust Lifting Parameterization

Most dust lifting parameterizations used in models use the model predicted wind or surface wind stress to determine whether dust lifting is active and further use the wind stress value to compute the dust flux. A subgrid lifting, usually associated with dust devils is often invoked as well (Newman et al. 2002; Kahre et al. 2006). Michaels (2006) recognized that the model predicted wind is, in the Reynolds averaging sense, the mean wind within the model grid cell and that dust devils are a member of a continuum of circulation sizes. Based on these two axioms, a new dust lifting model was developed.

On Earth, wind speed distributions are often well described by a Weibull probability distribution function. There is some evidence that the same distribution holds generally for Mars (Fenton an d Michaels 2010). This distribution has a tail at higher wind speeds, and the width of the distribution is related to the turbulence of the atmosphere. A purely laminar flow would have a distribution of zero width, and the winds would have a unit probability of being exactly the mean wind. Highly turbulent flows, such as those that are expected during a convective afternoon on Mars, would have a broad distribution with strong gusts and dust devils occupying the tail end of the distribution. Although the mean wind may not be of sufficient strength to lift dust, there is still a finite probability of winds exceeding a given lifting threshold.

If the model predicted surface shearing stress, as represented by $u_*$ is assumed to be the mean of a Weibull distribution, the fraction of values of $u_*$ exceeding a given threshold can be calculated if the width of the distribution is known. For simplicity, three possible distribution regimes are identified based upon the value of the bulk Richardson Number ($R_i$) in the lowest model layer. The first is a highly turbulent, convectively unstable and gusty regime corresponding to $R_i < 0$, the second is a moderately turbulent but stably stratified regime $0 \leq R_i \leq 0.03$, and the final is a laminar regime for $R_i > 0.03$. Note that although onset of turbulence is generally associated with $R_i = 0.25$, there is a hysteresis effect whereby turbulence can persist at higher values after onset. The width, $\gamma$, of the distributions for each regime is then set, with the largest width corresponding to the most turbulent case. Future versions of this scheme will utilize a continuous specification of widths functionally related to $R_i$.

Once the full Weibull distribution, $W$, is determined, the flux of dust, $F_d$, for a wind threshold above a critical lifting threshold, $u_t$, is normalized by the Weibull probability distribution function to generate the actual dust flux, $F_d^*$:

$$F_d^*(u_*) = W(u_*, \gamma) F_d(u_*) , \qquad u_* \geq u_t \qquad (1)$$

where

$$F_d = C_N (\rho_a \rho_d)^{0.5} u_*^3 (1 - \frac{u_t}{u_*})(1 + \frac{u_t^2}{u_*^2}), \qquad (2)$$

$C_N$ is an empirical constant, $\rho_a$ is atmospheric density, and $\rho_d$ is the density of a dust aerosol, following the work of Armstrong and Leovy (2005). Although $C_N$ is technically a bulk aerodynamic transfer coefficient, it must contain within information on the availability of surface dust. Thus, changing the value is equivalent to varying the dust lifting efficiency.

### Dust Aerosol Representation

In Rafkin (2009), only lifted dust was considered in the radiative transfer calculations. In reality, there exists a background dust concentration that is perturbed by local lifting, transport and sedimentation of dust. MRAMS now carries two distinct dust fields. The first is a background dust concentration and the second is a foreground perturbation dust field. The background dust concentration is static (i.e., no transport, sedimentation or surface sources/sinks) and may be initialized in any user-specified manner (including a zero background field). The background dust is always radiatively active. The perturbation dust field utilizes the microphysical bin model to track a user-specified number of dust mass bins subject to lifting, sedimentation and transport; the perturbation dust distribution is allowed to freely evolve. The perturbation field can also be initialized with any user-specified distribution.

The surface dust reservoir is explicitly tracked and tied directly to both the lifting and sedimentation of perturbation dust as a function of size. Dust lifted off the surface is assumed to follow a log-normal distribution with a user-specified mode (r≈1.5 μm). In most cases, the larger dust particles lifted from the surface fall rapidly to the surface leaving mostly dust of several microns and smaller in the atmosphere. However, a few large particles can contribute substantially to the total dust mass.

Often, the first few hours of model spin-up produce strong winds or transient circulations associated with the adjustment of the initial GCM conditions. These spurious circulations can lead to nonphysical lifting and perturbations of the dust field. Prevention of spurious lifting is achieved by disabling lifting within a user-specified period of time from initialization.

Both the background and perturbation dust fields can be independently radiatively active. All dust is assumed to have the properties of Hawaiian palagonite. The background dust size distribution follows a long-normal distribution (mode of r≈1.5 μm), with appropriate constants computed based on compositional properties and the assumed size distribution. The perturbation dust distribution optical properties are based on the instantaneous size distribution.

Although the legacy MRAMS radiation code based on the NASA Ames GCM (Haberle et al. 1993) is still available, the two-stream, correlated-k code must be used when perturbation dust is active; the legacy code makes assumptions about dust size that are not generally valid when the dust distribution is free to evolve. The radiative activity of dust can be switched on or off within simulations as desired. This ability is particularly valuable for assessing the radiative impact of lifted dust.

## Model Simulation Design

Ongoing MRAMS simulations conducted to assess the atmospheric environment at potential MSL landing sites are leveraged for this study. Here, results from Mawrth Valles landing site (~22.3N, 16.5W) simulations are presented and discussed. All simulations are configured identically with four grids, beginning with a super-hemispheric grid at 240 km horizontal grid spacing and decreasing by a factor of three on each successive grid down to 8.9 km on the fourth grid. Vertical grid spacing is ~10 m at the lowest level and is gradually stretched to a spacing not to exceed ~2 km. The top of the model is ~55 km.

Three scenarios with radiatively active lifted dust are examined through comparison to baseline simulations in which lifted dust is not radiatively active. The impact of dust on local and regional circulations and the role of WEIRD are then assessed. The first scenario is a global dust storm case, the second corresponds to a regional dust storm, and the third and final scenario investigates the impact of dust under more typical conditions.

To produce these three scenarios, the value of the lifting efficiency factor ($C_N$) is varied as a function of the model grid. For the global dust storm case, the value is set to a large number on all grids ($C_N = 9 \times 10^{-7}$), which results in large fluxes of dust into the atmosphere wherever lifting is permitted ($u_t = 14$ mN), and there is a rapid increase in atmospheric dust loading on all grids. In the second scenario, the efficiency factor is set to $9 \times 10^{-9}$ on all grids except for the fourth where the efficiency is increased to $9 \times 10^{-7}$. This has the result of producing a dustier fourth grid that mimics the development of local dust storm. In the final scenario, the efficiency is set to $9 \times 10^{-9}$ on all grids. This value has been found to produce reasonable dust distributions (i.e., perturbations of ~0.1 in opacity). A control case where lifted dust is radiatively passive is also conducted for each of these scenarios.

A major advantage to tuning $C_N$ to produce the different scenarios is that the dynamical field evolves in a completely self-consistent manner with the dust field. Dust is only lifted where the winds are of sufficient strength, and the circulation feels through radiative forcing the impact of the lifted dust. A disadvantage of this method is that the location and evolution of dust disturbances cannot be easily controlled; the disturbances form where the winds dictate and evolve according to the overall evolution of the circulation. Alternatives to specifying $C_N$ include initialization of the model with an incipient dust disturbance (as was done in Rafkin (2009)) or turning off lifting altogether and specifying the perturbation dust distribution as a function of time. The latter alternative allows for the precise control of when and where the dust disturbance appears and how it grows and decays. The disadvantage to this latter method is that the dust field is generally not consistent with the wind field; the winds respond to the specified dust, but the dust field is independent of the winds.

In all cases, a background dust field has been imposed in addition to whatever perturbation dust field may arise. The background dust field is identical to that used in the NASA Ames GCM simulations that are used to initialize and provide boundary conditions for MRAMS. The background dust prescription is based on TES observations of opacity and follows the Contrath-ν height profile (Conrath 1975), which is adjusted so that dust height varies seasonally with the subsolar point. The background prescription produces reasonably good agreement between the GCM and temperatures derived from TES radiance observations. Therefore, in the absence of perturbation dust lifting, the simulated Mars system behaves in a reasonably realistic manner. The addition of perturbation dust can be used to see how the climate system responds to physically consistent, local dust lifting.

## Numerical Simulation Results

If dust were a primarily passive tracer, all the simulations would evolve in a similar manner; lifted dust would have no impact on the circulation. As will be shown, this is not the case, indicating that lifted dust does alter the circulation not just on the global scale (as is previously known (Haberle et al. 1982, Murphy et al. 1995)), but also on the local and regional scale. Furthermore, it will be shown that the dust forcing is consistent with WEIRD.

### Local Impacts of Global Dust Changes

By setting the dust efficiency factor to a large value ($9 \times 10^{-7}$) on all grids, the entire model domain becomes increasingly dusty with time (Fig. 2). This mimics the growth of a global dust storm. Importantly, although the efficiency factor is large, the dust is lifted in a manner fully consistent with the model-predicted wind stress, and the dust is then allowed to circulate in a manner fully consistent with the dynamics. In the case with radiatively active lifted dust, the dust can alter the atmospheric heating profile and alter the dynamics.

Dust lifting is initiated not at the start of the simulation, but approximately 1.5 sols (mid-afternoon at Mawrth Valles) into the integration. This prevents any initial transients from lifting dust while the model spins up, as previously described.

The near-surface wind field and perturbation opacity at sunrise on the second sol, 15 hours after the start of dust lifting, is shown in Fig 3a for both the radiatively active and radiatively passive global dust storm case. Although the winds are generally similar in direction, the radiatively active dust case shows generally stronger winds. Also, the dust opacity fields have evolved differently, indicating that the radiatively active dust is having an influence on both the wind speeds and the atmospheric dust loading.

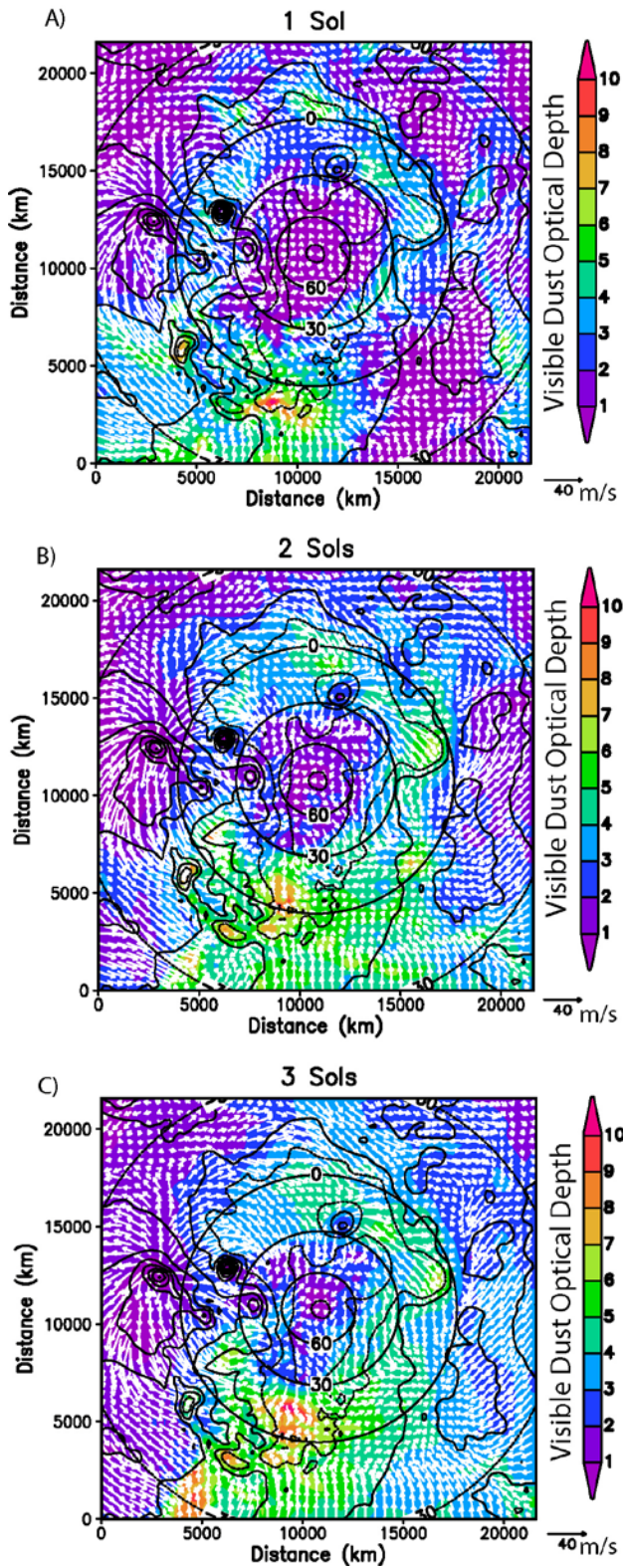

**Figure 2**. Dust and wind evolution in the global dust storm case. Lifting efficiency is set to 9x10-7 on all grids (fig 2.jpg).

Figure 3b shows the same simulations approximately 3.5 sols hours later during the afternoon. Dust opacity is 2 to 5 times greater in the radiatively active dust simulation. Wind speeds are slightly greater as well, but the wind directions are remarkably different. Whereas topography still exerts a strong influence in the radiatively passive simulation, as evidenced by strong convergence boundaries, channeling within Mawrth Valles, and more spatially variable winds overall, the radiatively active case has much more uniform winds that are much less strongly influenced by topography. Dust heating has become strong enough to overwhelm the thermally driven topographic circulations that typically dominate in nominal dust-loading scenarios. This effect is also seen in the pressure traces from a location near the proposed landing site (Fig. 4). As dust loading increases, the diurnal and semidiurnal tidal signatures strengthen while topographic pressure signals fade.

The topographic circulations are driven by temperature gradients that form when heating or cooling the surface. Elevated areas of topography will tend to warm compared to the surrounding free atmosphere during the day (*e.g.*, Ye et al. 1990). This induces buoyant vertical motion and through continuity results in upslope (anabatic) flows. The opposite occurs at night. Low level atmospheric dust reduces the formation of horizontal temperature gradients, because it causes the atmosphere surround the elevated topography to warm more than it would otherwise during the day.

In addition to the reduction in the intensity of the local topographic thermal circulations, the magnitude of the large-scale circulation associated with the diurnal and semidiurnal tide and large-scale dust-induced pressure gradient grows substantially. In the extreme case, the large-scale flow can overwhelm the local circulation with the result being near uniform winds in speed and direction.

Dust is clearly having a large impact on the local circulations and this has resulted in substantially different dust distributions. However, it is not clear whether the overall acceleration of the winds is related to the global increase in dustiness and an overall strengthening in large-scale winds, or whether the acceleration is being produced locally due to the local dust field. To differentiate between these two possibilities, the impact of a local rather than global dust disturbance on the local circulation is investigated.

## Local Impacts from Local Dust Storms

To assess the impact of local dust on the local circulation, a scenario with high dust lifting efficiency confined only to the fourth grid is analyzed. A local enhancement of dust centered within the fourth grid is produced in this manner. Cases with both radiatively active and radiatively passive dust are simulated. As in the global dust storm case, dust lifting is not activated until approximately 1.5 sols into the simulation (corresponding to mid-afternoon in the Mawrth Valles region). Also, like the global dust storm case, there is a prescribed, steady-state background dust field based on TES observations.

The perturbation dust opacity approximately three hours after sunrise (roughly 0900 local) on the second sol of the simulation is shown in Fig. 5 for both the control (radiatively passive dust) case and the radiatively active case on the third grid. Also shown is the difference in pressure (radiatively active minus control) and the wind vector difference. The third rather than fourth grids are analyzed here, because the impact of the dust within the fourth grid has the potential to perturb conditions beyond the fourth grid domain; it is important to identify any such impact. The perturbation dust opacities are generally below 0.5 in all locations with a few local areas exceeding 1.0. Wind speeds in both simulations are generally within a few m/s and directions are also similar. There are broad pressure differences between the two simulations resulting from differences in radiative cooling over the nighttime hours. Wind vector differences are greatest where these broad pressure differences are present (e.g., the western portion of the grid 3 domain). Also evident are locally intense negative pressure perturbations collocated with the highest dust opacities in the radiatively active case. Therefore, although there are some differences in the dust distribution between the two cases, dust has not substantially modified the circulation within the first two hours of sunrise; dust is behaving much like passive tracer, although not completely. There has been some divergence of solutions due to the lifted dust, and the heating within local regions of high dust loading are beginning to exhibit noticeable local pressure perturbations.

The situation changes dramatically by shortly after local noon (Fig. 6). While the dust optical depth in the control case is locally above 2.0 with broader regions closer to 1.0, the radiatively active case has peak optical depths over 5.0 with extensive regions over 2.0. Wind speeds are still 5-10 m/s in the control case but 10-15 m/s in the radiatively active. Peak pressure deficits have grown to 16 Pa, or ~2% of the total atmospheric mass. For reference, this corresponds to a ~20 mb low pressure deficit on Earth, which is a respectable storm system. Not surprisingly, the winds in the radiatively active simulation have responded to the pressure gradient with a predominantly convergent flow; little evidence for rotation is present. Both simulations have a line of convergence separating northerly and southerly winds, but it is stronger in the radiatively active case. The strongest lifting is along this convergence boundary.

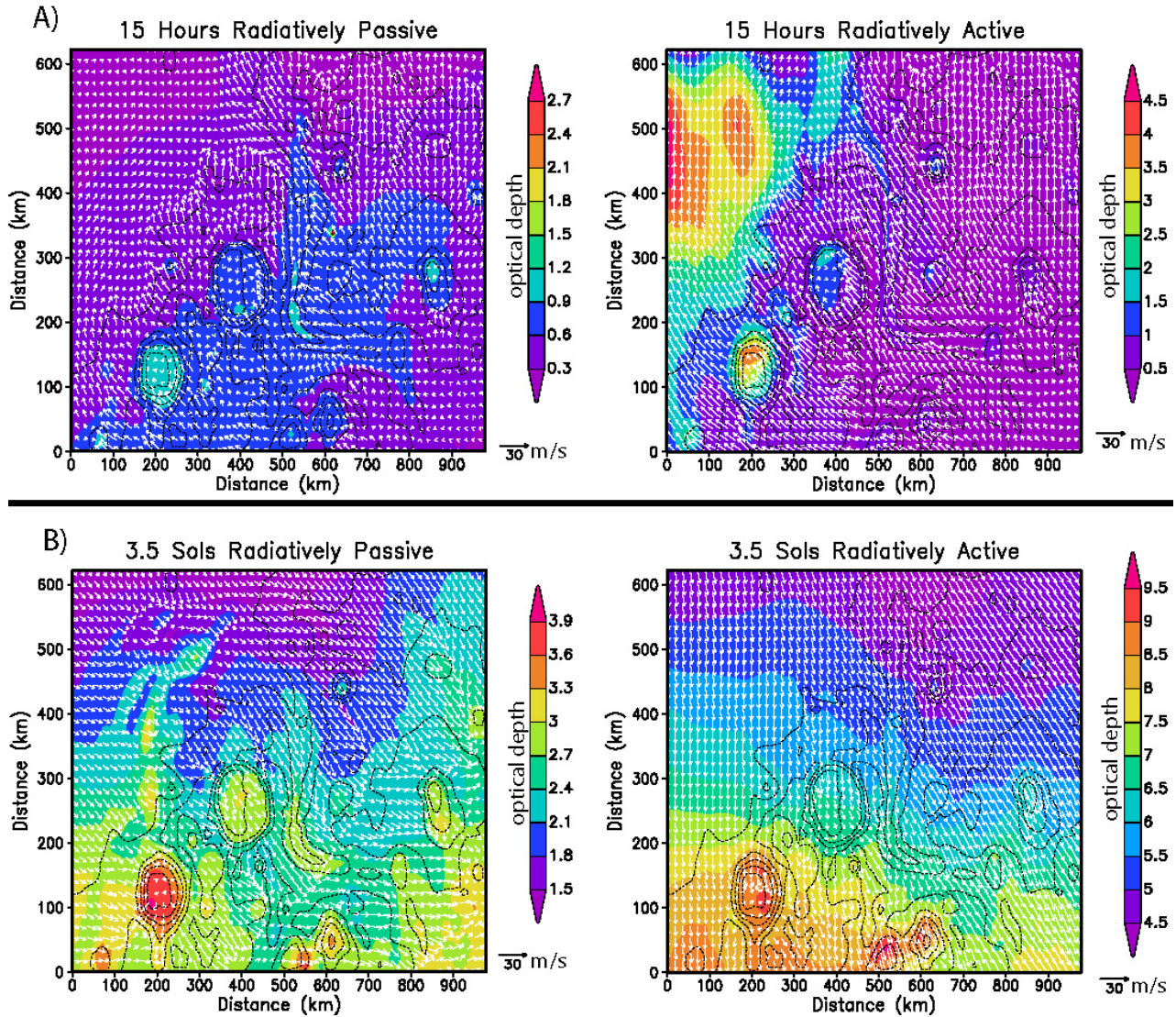

**Figure 3**. Global dust on grid 4 for the radiatively passive (control) and radiatively active foreground dust.  A) 15 hours after dust lifting, and B) 3.5 Sols later.  Dust opacity is shaded.  Topography is contoured (fig3.jpg).

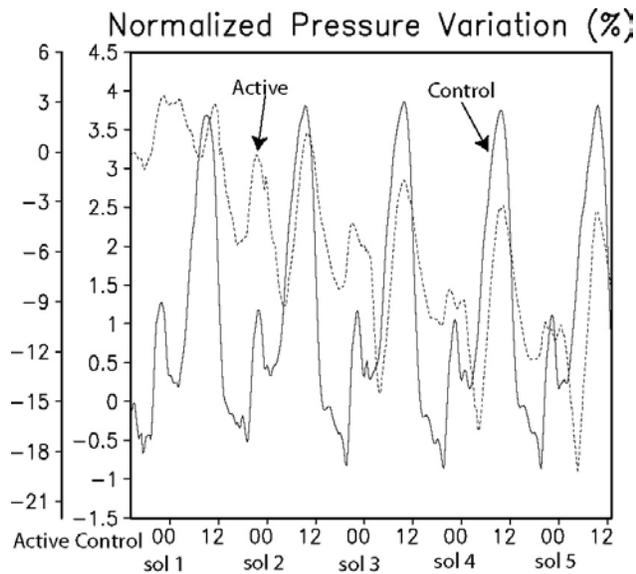

**Figure 4**. Global dust pressure trace on grid 4 normalized by the initial pressure value at the start of the simulation. Leftmost axis is pressure change (%) corresponding to active dust case (dashed). Rightmost axis is pressure change (%) corresponding to passive dust case (solid). The radiatively passive perturbation dust case (solid) exhibits a typical diurnal/semi-diurnal signal, with noticeable higher frequency perturbations, particularly during the daylight hours. The overall pressure cycle is stable. In the radiatively active perturbation dust case (dashed), there is a noticeable downward trend in pressure as the atmospheric column heats. The diurnal and semidiurnal amplitudes amplify substantially while the higher frequency, topographically forced, signals are overwhelmed (fig4.jpg).

By about 1500 local time, the system in the radiatively active case reaches maturity based on dust optical depth (Fig. 7). Winds are now nearly double that of the control case (a 4x increase in kinetic energy) and two well-defined cores of dust and pressure deficit are present. Both exhibit rotational wind tendencies around the core, which are located on the preexisting convergence boundary. The rotation would be more obvious if the background wind were removed; the perturbation wind field calculation partially compensates for the influence of background winds, and rotation is clearly evident. Had much of the surface not been depleted of dust by this time, it is possible that the system would have been even more intense. Dust lifting and wind speeds begin to decrease after this time due to the loss of solar heating;

however the convergence boundaries, cyclonic circulations and collocated dust maxima persist for many hours into the evening.

It is worthwhile to look into the structure of this system in greater detail in both time and space. Grid 4 (~8.9 km horizontal spacing) contains the core of the western disturbance during the mature phase, which will therefore be the focus of the subsequent description. The eastern circulation is at the western boundary of the 4th grid and is not completely contained within, but can still be partially identified.

The western system propagates from west to east over the Mawrth Valles region during the course of the afternoon. What appeared as a single convergence boundary on the 3rd grid (e.g., Fig 7) is now clearly an initial complex of intersecting boundaries (Fig. 8a). Dust maxima are located along the convergence boundaries as a result of the strong winds accelerating toward these boundaries, and due to the focusing effect of the convergent flow on the dust. Peak dust values are nearly collocated with the intersection of boundaries. Toward the center of the ring of convergence boundaries and dust, the winds are at a minimum and produce a net divergence as the flow spreads outward toward each of the boundaries. The dust optical depth is at a local minimum at the center of the divergent circulation.

This mesoscale organization of the circulations resembles open cell microscale convective circulations that occur within a convective boundary layer (e.g., Michaels and Rafkin, 2004), but at a much larger scale. Like the microscale circulations, the strongest winds are located along the convergence boundaries and these locations also tend to be the dustiest places. Dust devils and convective vortices are most common at the intersection of microscale convective eddy boundaries in a manner analogous to what is seen here at the mesoscale. The same mesoscale organization is seen in the control case, so dust is not the root cause. The organization is instead an emergent property of the system. However, the dust clearly acts to amplify the strength of the circulation, because the dust opacity and winds are greater in the radiatively active case.

After three more hours (roughly 1500 local), additional mesoscale organization has taken place. Many of the previous convergence boundaries have faded and one disturbance in the western half of the domain has emerged (Fig. 8b) as the dominant disturbance. This type of organization is not dissimilar in appearance from the mesoscale organization of individual thunderstorms into a mesoscale convective complex or a tropical cyclone on Earth. The reason for this organization in the simulation is unclear, but given the suggested thermodynamic and dynamic similarities of intense dust storm systems to tropical cyclones on Earth (Rafkin, 2009), the organization mechanisms may also be similar.

An hour later (Fig. 8c), the convergence boundaries continue to consolidate into a single dynamical entity. The boundaries describe a roughly cyclonic circulation around the disturbance. Again, the mean wind flow helps to obscure the storm-relative circulation. Dust optical depth reaches its maximum value of ~5.5 at this time and is loosely organized into a ring surrounding a low-opacity region near the center of the circulation where the convergence boundaries intersect. Morphologically, this is not dissimilar to a hurricane eye.

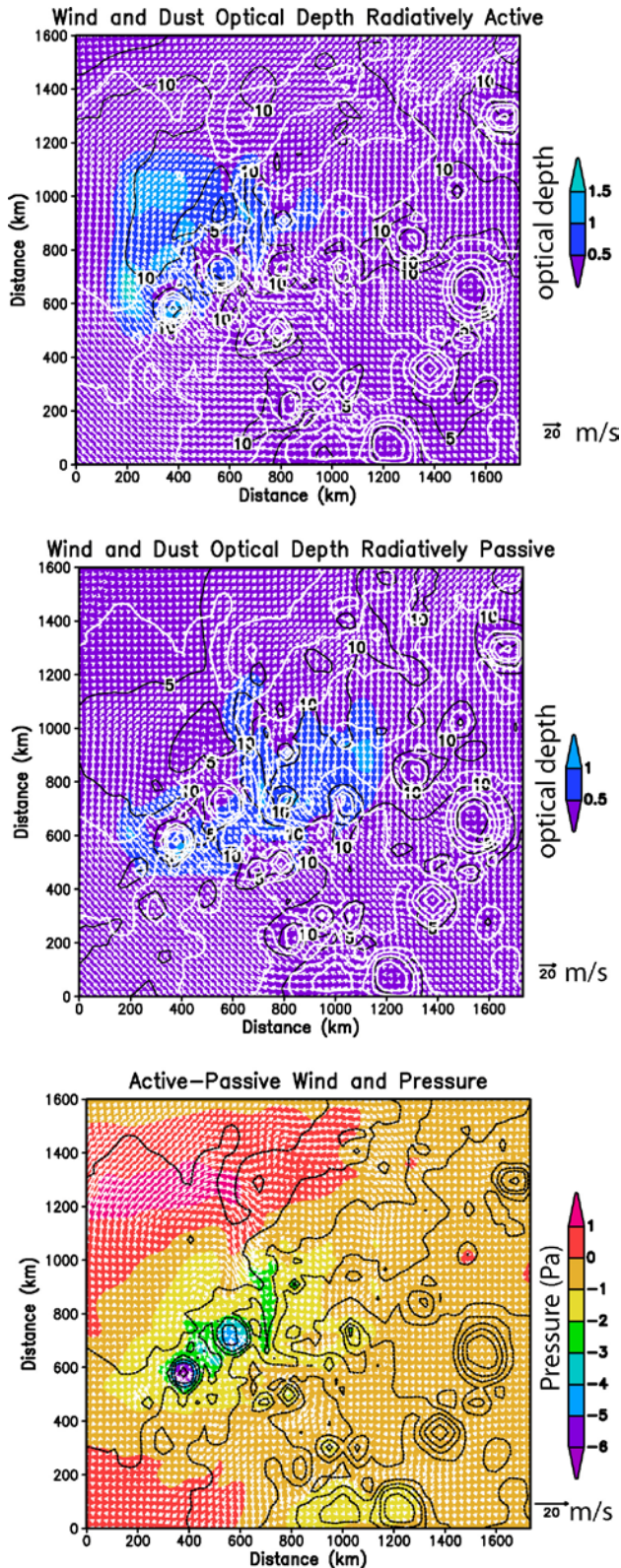

**Figure 5.** Winds and dust opacity for radiatively active (top) and passive (middle) on grid 3. Bottom: radiatively active pressure minus passive (shaded in Pa) and opacity active minus opacity passive (contoured) on grid 3 (fig5.jpg).

The system turns more to the southeast and the peak dust optical depth falls to ~4 one hour later (Fig. 8d), and to under 3 another hour after that (not shown). Although the dust opacity is falling, a cyclonic perturbation is clearly evident having now had sufficient time to respond dynamically to the warm core pressure deficit; note the strong curvature of the winds in the high dust optical region in Fig. 8d. By three hours from maximum development, the only notable dust perturbation in the grid 4 domain is associated with the decaying system as it nears the southern domain boundary (not shown). By nightfall, the nocturnal inversion begins to develop, and downslope (katabatic) flows originating from the higher terrain in the south and from ridges and crater rims develop and begin to overwhelm the vortex (not shown).

A vertical cross-section from the fourth model grid cutting N-S through the center of the storm center as indicated in Fig. 8c is shown in Fig. 9. The vertical wind structure shows 15+ km deep updrafts along the primary storm convergence boundaries. Above this are strong gravity waves propagating out and away from the disturbance. Dust mixing ratio is maximized in the updraft cores, and the development of an elevated dust layer can be seen in well defined outflow away from the storm center. There is a horizontal jet in the dusty outflow aloft.

Subgrid scale turbulent kinetic energy is a good indicator of the depth of the convective mixed layer. The dust storm updrafts tower above the surrounding boundary layer, which is approximately 6 km deep. Dust is injected well above the boundary layer to produce the elevated maximum in dust mixing ratio. Deep non-local transport, such as exhibited within the dust storm is an efficient way of producing high-altitude layers of dust (or water, methane or any other substance with a source in the boundary layer) that have most recently been observed by MCS (Mcleese et al. 2010; Heavens et al. 2011).

The structure and evolution of the dust storm in the radiatively active dust case should be compared to the control case where high dust optical depths are present, but where the lifted dust can have no thermodynamic or kinematic forcing. Like the radiatively active case, there are very distinct mesoscale convergence boundaries. The major difference, however, is that in the control case, the majority of convergence boundaries are anchored to the topographic relief (Fig. 10). As previously noted, the dust optical depth and the winds in the control are generally a factor of two smaller than the radiatively active case. For the most part, maxima in dust are collocated with the topographic convergence boundaries, which is to say that the dust lifting sources are anchored to the topography. Notably, there are no convergence boundaries in the northwest corner of the grid where the topography is flat, but where the dust storm formed in the radiatively active case. Therefore, a comparison of the radiatively active case to the control case suggests that topographic circulations play an important role in focusing dust lifting, and the radiative forcing from this dust does influence and modulate the circulation. At the same time, the comparison also shows that dynamically active dust systems can develop from dust perturbations that are not a product of topographic dust lifting. This is what was found in Rafkin (2009), and is further consistent with the expectation that topography may actually disrupt the formation of a dynamical storm system.

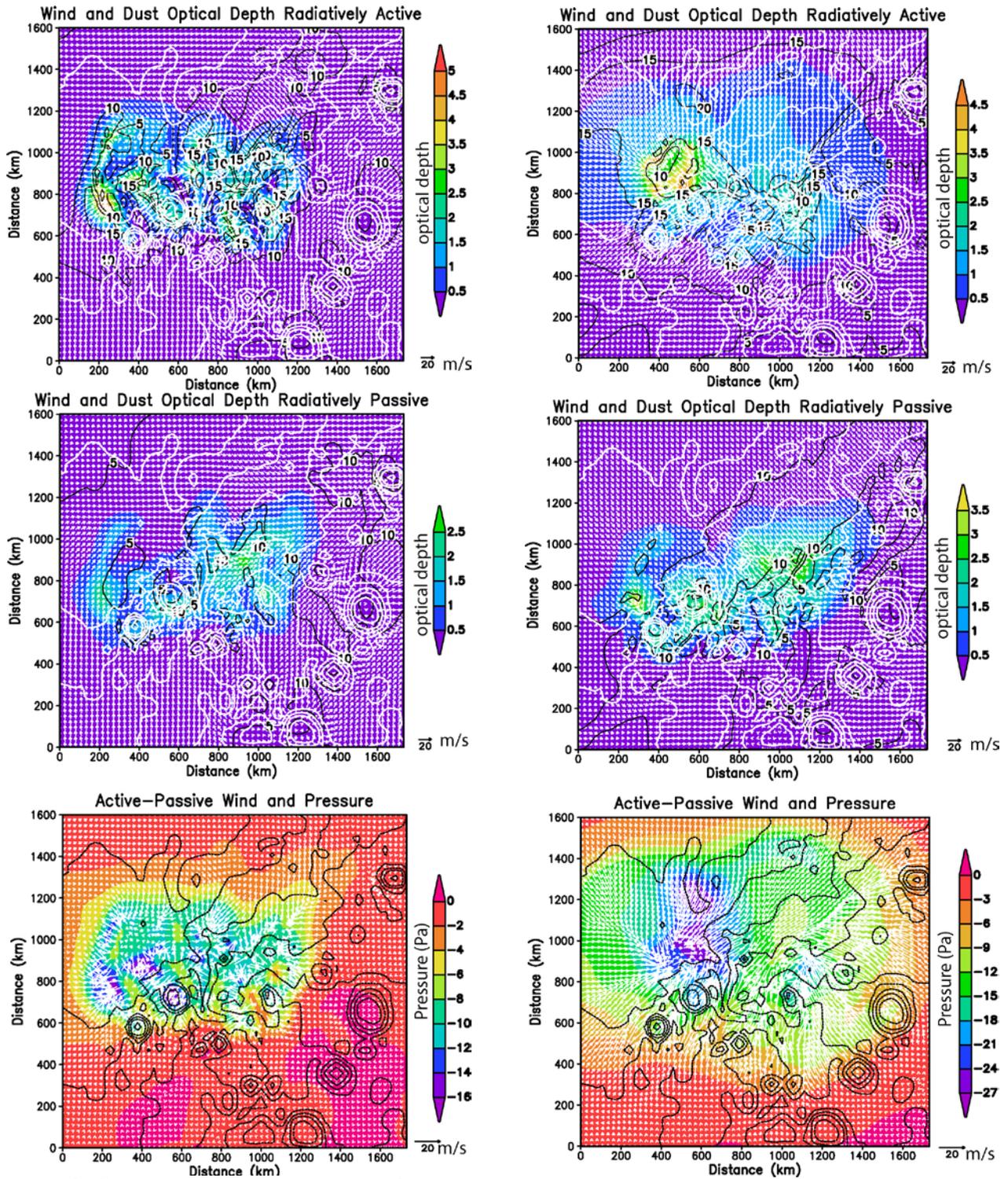

**Figure 6**. Same as Fig. 5, but approximately 6 hours after sunrise on the second sol. Top panels are the same as Figure 8. Lower panel is the difference in pressure and vector winds. Dust is having a profound impact on the circulation with enhanced convergent flow into the low pressure center of the storm. Peak optical depth in the radiatively active dust case is nearly double that of the control case and the circulation is dustier overall. This indicates an active feedback process between the lifted dust, radiation, and dynamics (fig6.jpg).

**Figure 7**. Comparison of radiatively active and control case for the local dust storm approximately 9 hours after sunrise on the second sol. Panels are as in Fig. 6. Winds in the radiatively active dust case are now showing cyclonic rotation around the strong pressure deficits, particularly in the core of the eastern disturbance. Note the influence of the storm on the winds beyond the immediate vicinity of the active lifting areas in the radiatively active case (fig7.jpg).

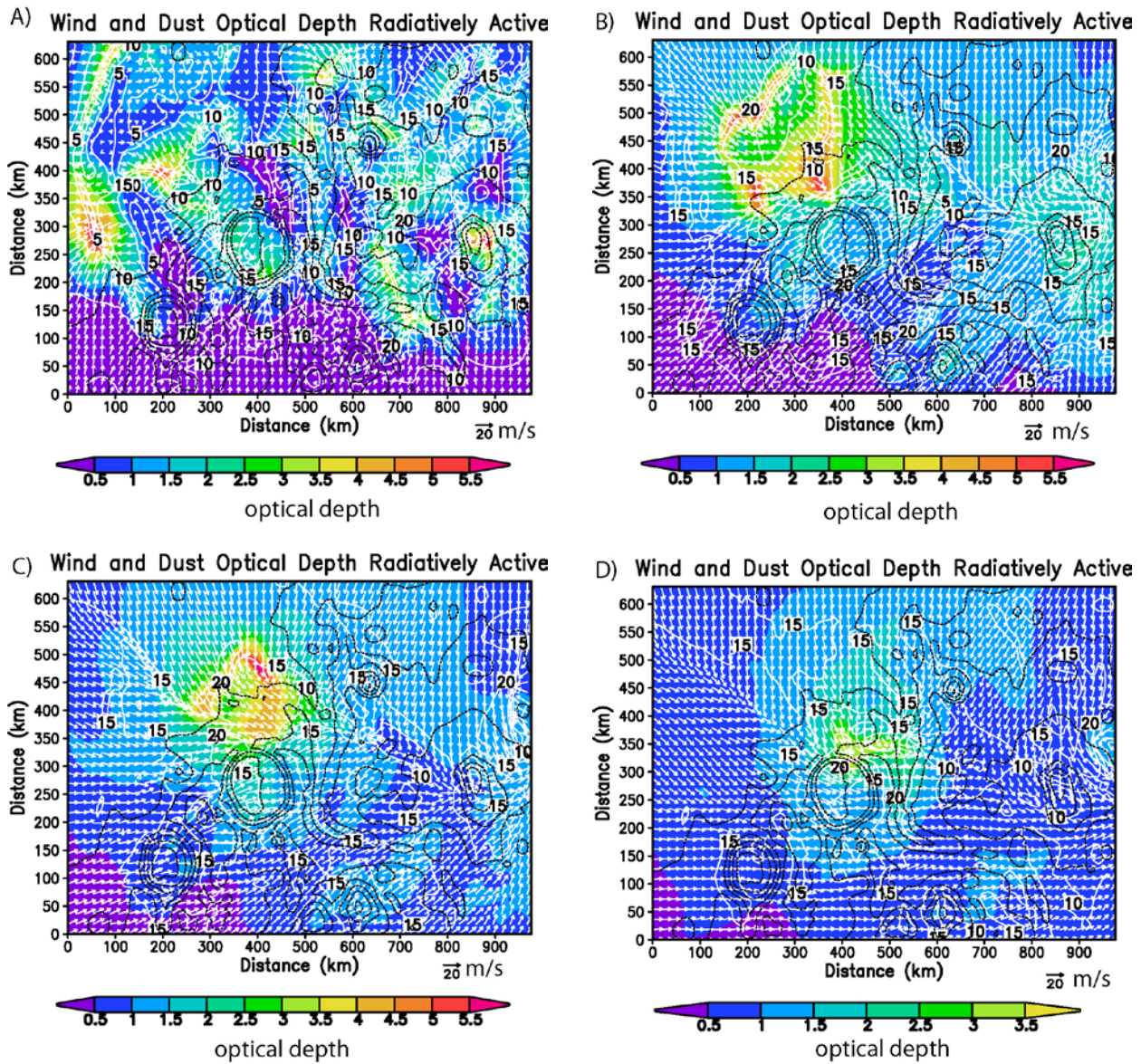

**Figure 8.** Grid 4. Dust disturbance as simulated on the 4th grid. Top left noon (t=68 same as Fig. 6), top right t=1500 (t=77 same as Fig. 7); bottom left t=1600 (t=80). The red line indicates the location of the vertical cross-section in Figure 9; bottom right t=1700 (t=83). Dust opacity is shaded (fig8a.jpg, fig8b.jpg, fig8c.jpg, fig8d.jpg).

In the vertical, there are dramatic differences between the radiatively active and the control cases. The control case has strong vertical updrafts along convergence boundaries in which dust is lifted, but the circulations are much less vertically extended. Nonetheless, updrafts along crater rims and valley walls can rise many kilometers above the surrounding boundary layer (Fig. 11), and they do so without the aid of radiative heating from the foreground dust. Not surprisingly, these circulations also produce elevated dust layers, but they are not nearly as well defined as in the radiatively active case. In the next section the impact of more typical dust loading on the local atmospheric circulation is explored.

## Local Impacts from Typical Dust Loading

Large injections of dust clearly have large impacts on the local circulation. Further, even in the absence of radiative heating of the lifted dust, topographically driven circulations are capable of transporting dust into the free atmosphere. To better assess the role that more typical values of dust loading have on circulations, including any impact such dust may have on the dust transport within local topographically-driven circulations, a third set of numerical experiments are conducted. In this final set of experiments, dust lifting efficiency is set to low to moderate values of $9 \times 10^{-8}$, $9 \times 10^{-9}$, or $9 \times 10^{-10}$ and the lifted dust is radiatively active. At a value of $9 \times 10^{-10}$, dust optical depth perturbations of up to ~0.01 are found. Peak optical depth scales approximately linearly with an increase in lifting efficiency so that peak values of 0.1 and 1.0 correspond to values of $9 \times 10^{-9}$ and $9 \times 10^{-8}$, respectively.

For the case of efficiency set to $9 \times 10^{-10}$, the differences between the radiatively active and radiatively passive simulations are trivial. Wind speeds rarely differ by more than a fraction of a m/s, and temperatures differ by no more than a fraction of a Kelvin, and both deviations are extremely localized. Dust fields are nearly identical. Small, localized changes in dust opacity on the order of 0.01 or less have effectively no meaningful forcing at the local scale; dust behaves very nearly like a tracer quantity. Dust is still transported vertically in strong topographic circulations and elevated dust layers are produced (identical in structure to those seen in Fig. 11, but with less dust due to the smaller lifting efficiency), but there is insignificant feedback between the dust and circulations that lift the dust.

A lifting efficiency of $9 \times 10^{-9}$ is perhaps the most realistic. Peak dust perturbations are of the order 0.1, but perturbations are typically closer to 0.01. Locations with the higher amount of dust loading are now able to drive more significant changes. Wind speeds now differ by up to a few m/s, temperature by a few Kelvins, and pressure by up to 0.1%. Although there is a quantifiable feedback, it is not sufficient to generate a dust storm in any dynamical or optical sense.

Although the lifted dust concentrations do not produce a rapidly growing disturbance with $C_N=9 \times 10^{-9}$, the dust does exert an influence, which can be seen in the vertical transport. With radiatively active dust, there is a tendency for the dust to be lofted somewhat higher than in the radiatively passive case (or the case of $C_N=9 \times 10^{-10}$), and more pronounced elevated dust layers form (Fig. 12). Therefore, the finding that topographically forced thermal circulations loft dust and produce elevated dust layers in the radiatively passive case also holds under scenarios with realistic dust lifting and radiatively active dust. Although these dust layers are not as deep or as well defined as those of the simulated dust storm, topographic thermal circulations are ubiquitous and persistent sol after sol. Therefore, their impact may be extraordinarily large and could play a key role in the overall global dust cycle; these systems may provide the key transport mechanism that transports dust out of the boundary layer and into the free atmosphere so that the global atmospheric dust load is maintained.

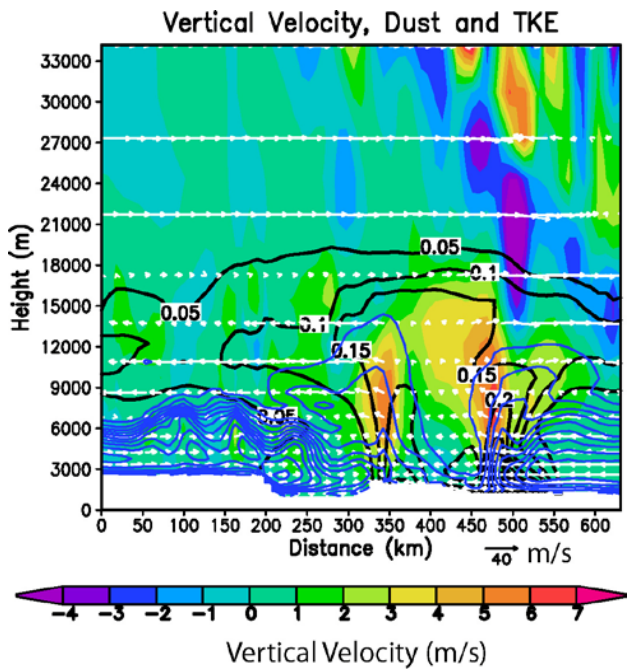

**Figure 9.** Grid 4 vertical cross-section along the line shown in Fig8c. Vertical velocity (m/s) is shaded, Dust mixing ratio (g/kg) is contoured in black, and sub grid scale TKE is contoured in blue in order to show the convective mixed layer. Updrafts in the storm system penetrate through the boundary layer to produce deep local transport of dust. An elevated dust layer extends outward above the free atmosphere. Strong gravity waves propagate upward and away from the disturbance (fig 9.jpg).

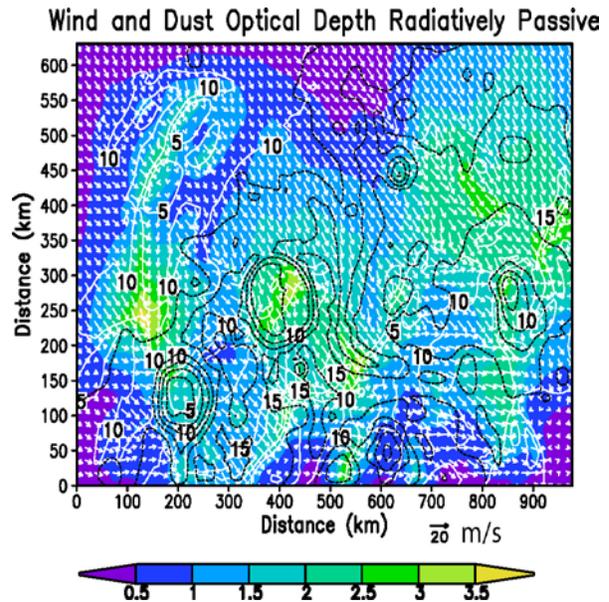

**Figure 10.** Near-surface wind pattern and dust opacity (shaded) for the radiatively passive case at the same time as that in Fig. 9. Convergence boundaries and dust lifting are focused by topography (contoured) (fig10.jpg).

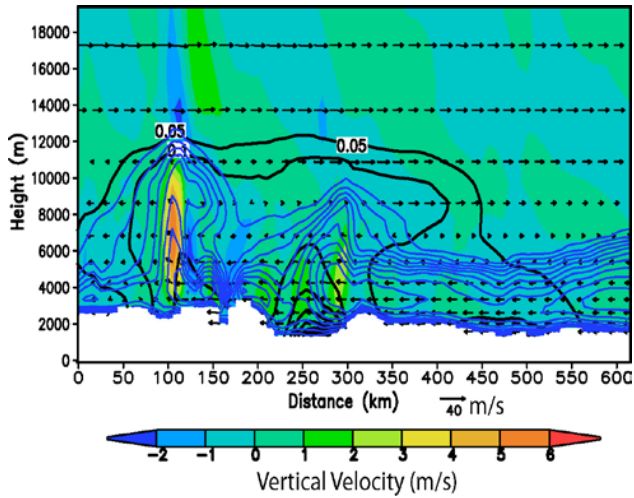

**Figure 11**. Grid 4 vertical cross-section through a strong thermal topographic circulation (x=50 Fig. 10) in the radiatively passive case. Vertical velocity (m/s) is shaded, Dust mixing ratio (g/kg) is contoured in black, and sub-grid scale TKE is contoured in blue in order to show the of the convective mixed layer. Like in the dust storm (Fig. 9), updrafts penetrate through the boundary layer to produce deep local transport of dust. An elevated dust layer extends outward above the free atmosphere, but it is not nearly as well defined as in the dust storm. Strong gravity waves propagate upward and away from the disturbance. Note the vertical scale difference between Fig. 9 (fig11.jpg).

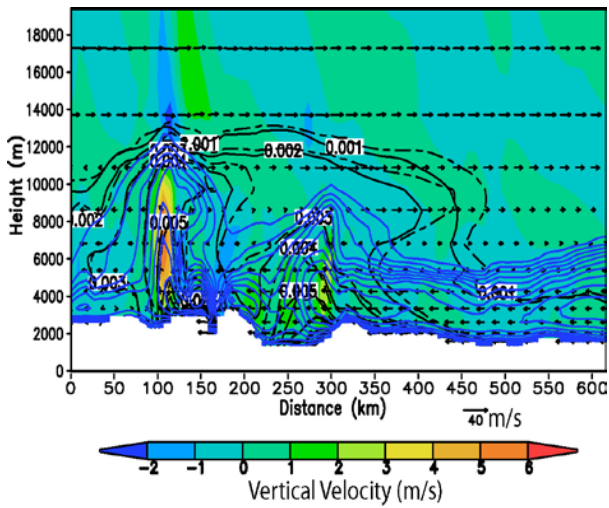

**Figure 12**. Same as Fig. 11, but for dust efficiency set to 9x10-9 on all grids. Dust mixing ration for the radiatively active case is solid, passive is dashed. The active case has slightly deeper dust and slightly more extensive elevated dust layer. Vertical velocity and TKE are shown in the background for the passive case (fig12.jpg).

## Radiative Impacts of Concentrated Low Altitude Dust

The heating of the atmosphere by radiatively active dust is not surprising; the warming of the atmospheric column by dust has been known for some time. However, what has typically been assumed is that there is a tendency to cool the near-surface layers due to attenuation of solar radiation at higher altitudes, and there is heating aloft where the solar flux is absorbed. In the case where dust is well mixed to relatively high altitudes, as is assumed in a typical Contrath-v profile, this scenario is accurate and produces an increasingly stable atmosphere that tends toward isothermal conditions (Pollack et al. 1982).

In the early and mature phases of active dust lifting, the assumption that dust is distributed deeply throughout the atmosphere is violated. Instead, large amounts of dust are concentrated within the lowest scale height even though the dust optical depth may be greater than 5. This vertical distribution of dust has a profound effect on the heating and temperature profile. When dust is concentrated in the lower atmosphere, solar absorption is concentrated in the lower atmosphere rather than being spread through a deep atmospheric column. This results in very large heating rates in the lower atmosphere. Of course, the solar heating is greatly reduced right near the surface after passing through an optically thick dust layer, but the layer itself is strongly heated by the solar absorption. Furthermore, this absorption occurs in the low levels where the air is already generally relatively warmer than it is aloft. Solar heating of the low level dusty layer causes a direct increase in temperature that dramatically increases the emitted infrared emission. This energy is emitted upward, where some is absorbed on its way to space. Energy is also emitted downward where some is absorbed by the dusty air and the rest by the surface, which also then heats. The surface emits infrared radiation upward, where even more energy can be absorbed by the low level dusty air. In short, high concentrations of low level dust produce a very effective greenhouse effect, which causes the temperature of the low level air to rise dramatically above that which is found in a more typical situation. This occurs despite the large attenuation of solar energy. It is not until optical depths are greater than 5 that solar absorption becomes so strong and localized aloft that a surface inversion begins to form.

Fig. 13 shows the surface radiation budget and atmospheric heating rates obtained from a column model for a few dusty radiatively active cases and one with only typical background dust. Note the large attenuation of solar energy in the very dusty cases with the intensity of heating increasing as dust is confined to lower altitudes. Solar energy is being absorbed in a shallow layer near the surface when dust is confined to low levels. The solar heating results in an increase in temperature and a dramatic increase in infrared energy emission compared with a typical background dust environment.

It has been suggested that the reduction of solar heating near the surface by dust storms may cause the systems to decay due to a loss of surface insolation (Zurek et al. 1982). This is not the case in the early and mature stages of an active dust lifting area. The increased dust loading does not reduce the heating of the ground and low-level air. The opposite occurs.

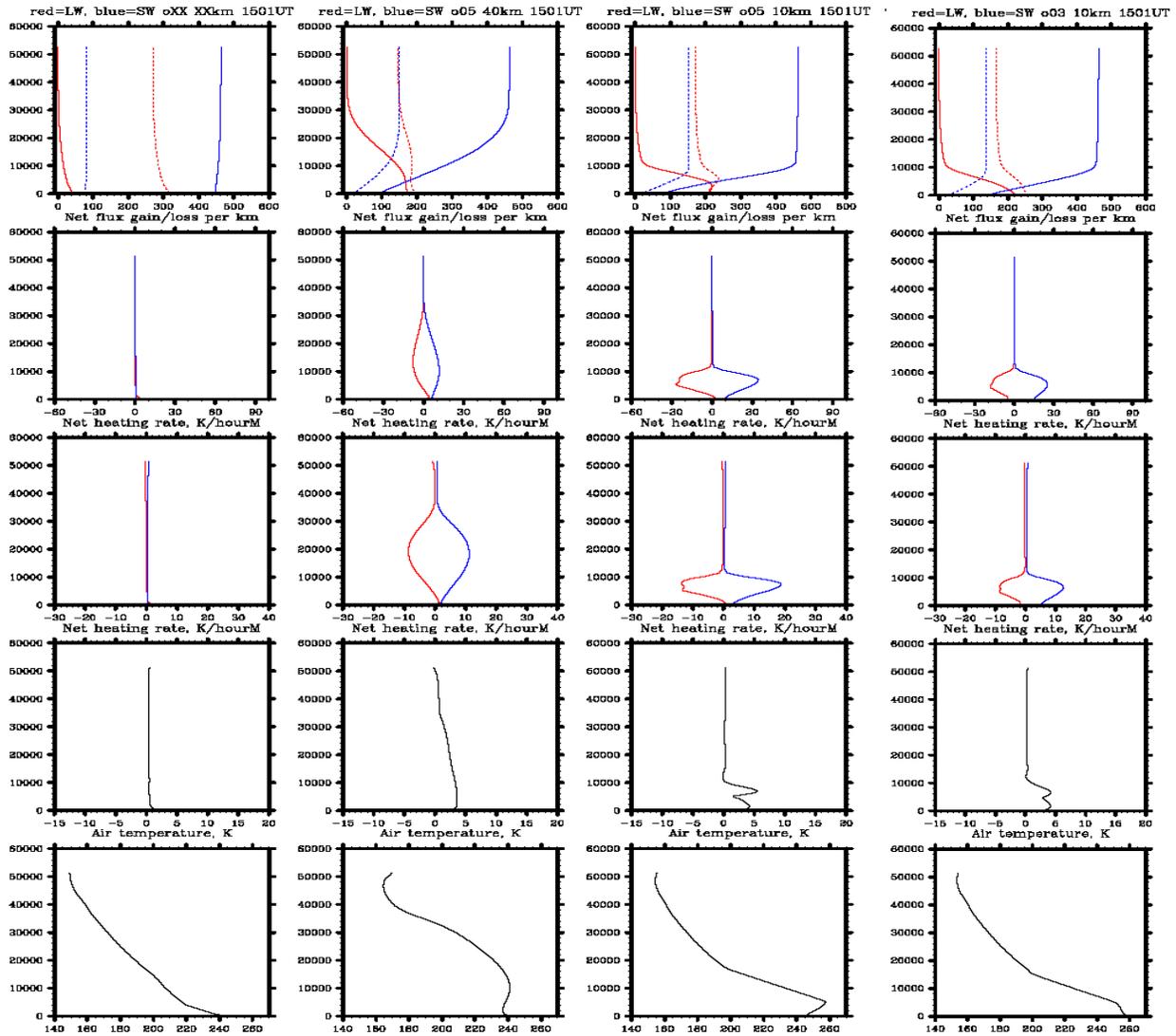

**Figure 13**. Radiative fluxes, heating rates and temperatures from the column model for various dust profiles. 1st column: Conrath-v dust profile with typical optical depth (~0.3). 2nd Column: Optical depth of 5 distributed over 40 km following Conrath-v like profile. 3rd column: Optical depth 5 with dust in the lowest 10 km. 4th Column: Optical depth 3 with dust in the lowest 10 km. Top row: solar (blue) and IR (red) downward (dashed) and upward (solid). 2nd row net flux / km. 3rd row heating rate K/hour for visible and IR. 4th row total net heating rate. Last row: temperature profile (fig13.jpg).

The dust increases the low-level temperature causing a further reduction in pressure, a further acceleration of the winds, and a further increase in dust lifting. In the very dusty simulations, it is not atypical to find a 20+ K increase in the temperature of the lowest atmospheric layer. It is not until dust is mixed very high into the atmosphere and opacities are greater than ~5 that the radiative forcing tends to stabilize the atmosphere by cooling the low levels and warming aloft. Since the dust storm updrafts do not extend to 30 or 40 km, the deep mixing of the dust to these altitudes must be accomplished by the more sluggish large-scale circulation. Thus, the timescale on which the dust will have a negative radiative impact of storm growth is greater than the time scale of growth from onset to maturity. Only storms that are active for extended periods of time (perhaps a sol or greater) will be affected by the deep redistribution of dust by large scale circulations.

Above the shallow dust layer, the atmosphere has a very steep lapse rate. If dust storms operate as Carnot engines, then this thermal structure would be conducive for further development. Steep lapse rates result in the optimal situation where warm air flows into the system and the system exhausts air at a cold temperature. Deeper dust distributions result in a more isothermal structure, which would tend to inhibit growth.

## Summary and Conclusion

Atmospheric dust loading is known to strongly influence the temperature and kinematic structure of the atmosphere due to its radiative properties. Until recently, the focus of dust impacts has been on the global scale. Here, the impact of locally lifted dust on the local and regional scale has been investigated using simulations of the Mawrth Valles region. Dust lifting efficiency was varied to produce varied levels of lifted dust, and that lifted dust could be either radiatively active or radiatively passive. Through comparison between simulations with the radiatively active and passive dust, the impact of radiatively active dust on the atmospheric circulation was assessed. For the cases where dust lifting efficiency was large, large amounts of dust were placed into the atmosphere at either the global scale or at the regional scale, and in both cases the atmospheric circulation was found to respond strongly to the perturbed radiative forcing.

WEIRD is a viable hypothesis to explain the growth and maintenance of both local and regional dust storms and of active lifting centers that ultimately lead to larger-scale dust events. Generally, the addition of radiatively active dust strengthens the atmospheric circulation that lifts the dust. Therefore, there is a positive radiative dynamic feedback whereby dust increases the near-surface winds so that more dust is injected into the atmosphere.

Although no hurricane-like systems like those seen in Rafkin (2009) developed in the realistic simulations, many of the characteristics identified in the idealized storms were found in the more realistic situations. However, there was never an expectation that dust hurricanes would form on Mars. Rather, it was suspected that the WEIRD processes operating to produce the highly organized systems in the idealized cases might also be operating under realistic conditions. This appears to be the case. In particular, the lifted dust created a warm core low pressure system with a tendency for low-level cyclonic rotation and anticyclonic rotation at the top of the system. In addition, convergence boundaries with enhanced vertical velocity, winds, and dust lifting were produced, although these were not nearly as well organized as the spiral arm bands simulation in Rafkin (2009).

If some dust storms do possess some degree of organization, the question of why such structures are not easily seen in orbital imagery needs to be addressed. First, it's possible that such structures are in the imagery, but no dedicated investigation has been undertaken to identify them. Second, unlike hurricane structures that are revealed by water clouds, atmospheric dust concentrations are not sensitive to the sign of the vertical motion (i.e., upward or downward). Hurricane structures are obvious, because clouds form where air is rising and dissipate when it descends. Sharp gradients between rising and sinking air in hurricanes show up as well-defined cloud boundaries, such as eye walls and spiral arm bands. The simulations also show sharp gradients in vertical motion, but the dust field does not respond like water. If dust did evaporate as the air descended and warmed, structures would be much easier to identify. Third, once a large optical depth is reached, the dust appears opaque and the surface is obscured. Without a contrasting background, it is difficult if not impossible to see gradients in otherwise high values of optical depth. For example, the disturbance in Fig. 8d would likely appear as a very dusty blob, whereas contour lines and color shading at quantized intevals makes the structure visible. Related to the obscuration of details by thick dust is the obscuration by the elevated dust layers that are produced. An optically thick dust layer above the boundary layer is in the outflow region of the storm, which behaves dynamically different from the lower region where much of the structure may be found. Not only can this elevated dust layer obscure what lies below, but there may be little dynamical structure in the elevated layer to which the dust can respond. Finally, it is possible and even likely that most dust disturbances, while feeding off WEIRD instabilities, are in an environment that is much too disruptive to allow the formation of a highly structure system. Nevertheless, a lack of visible structure does not mean that WEIRD or other feedback processes are not operating.

Boundary layer turbulence may be capable of mixing dust through the boundary layer, but these eddies cannot effectively mix dust out of the boundary layer. Some other mechanism(s) must be operating in order to maintain the global atmospheric dust load at altitude, and the non-local deep transport and venting associated with dust storms and with topographic circulations are excellent candidates. Rafkin et al. (2002) identified similar transport associated with Arsia Mons and likely other large volcanic features, but it now appears that smaller, more mundane topographic features are also capable of lofting dust. Once dust is in the free atmosphere, larger-scale circulations can transport and mix the dust globally and to higher altitudes in order to produce the observed background dust field in the atmosphere of Mars.

The dust within updrafts plays an important role in increasing the effectiveness of non-local deep transport. Radiatively passive dust can still be injected by strong, vertical thermally forced topographic circulations into the free atmosphere above the convective boundary layer, but the altitude of penetration is lower and the production of well defined elevated dust layers is more limited compared to the case with radiatively active dust. Therefore, dust plays an important role in influencing its own transport. Any tracer species present in the boundary layer with the dust (e.g., water and chemical species, possibly including $CH_4$) will be transported in a manner similar to the dust. Vertical transport of these quantities will be deeper when the updrafts are dusty.

## Directory of supporting data

## Acknowledgements

The numerical simulation work was supported by the Mars Science Laboratory (MSL) Project at the Jet Propulsion Laboratory. Coauthors at the University of Colorado were supported under a subcontract of this parent MSL contract. Analysis of the model data was unsupported and was conducted by the author at his own expense.

## References

Armstrong, J. C., and C. B. Leovy (2005) "Long term erosion on Mars" Icarus 176, 57-74, doi:10.1016/j.icarus.2005.01.005.

Conrath, B. J. (1975) "Thermal structure of the Martian atmosphere during the dissipation of the dust storm of 1971" Icarus, 24, 34-46, doi:10.1016/0019-1035(75)90156-6.

Emanuel, K. A. (1991) "The theory of hurricanes" Annual Review of Fluid Mechanics 23, 179–196, doi:10.1146/annurev.fl.23.010191.001143.

Fenton, L. K. and T. I. Michaels (2010) "Characterizing the sensitivity of daytime turbulent activity on Mars with the MRAMS LES: Early results" Mars, 5, doi:10.1555/mars.2010.0007.

Haberle, R. M., C. B. Leovy, and J. B. Pollack (1982) "Some effects of global dust storms on the atmospheric circulation of Mars" Icarus, 50, 322-367, doi:10.1016/0019-1035(82)90129-4.

Haberle, R. M., J. B. Pollack, J. R. Barnes, R. W. Zurek, C. B. Leovy, J. R. Murphy, H. Lee, and J. Schaeffer (1993) "Mars atmospheric dynamics as simulated by the NASA Ames general circulation model 1. The zonal-mean circulation" Journal of Geophysical Research, 98, E2, 3093-3123, doi:10.1029/92je02946.

Heavens, N. G., D. J. McCleese, M. I. Richardson, D. M. Kass, A. Kleinböh, J. T. Schofield (2011) "Structure and dynamics of the Martian lower and middle atmosphere as observed by the Mars Climate Sounder: 2. Implications of the thermal structure and aerosol distributions for the mean meridional circulation" Journal of Geophysical Research, 116, E1, doi:10.1029/2010je003713.

Kahre, M. A., J. R. Murphy, and R. M. Haberle (2006) "Modelling the Martian dust cycle and surface dust reservoirs with the NASA Ames general circulation model" Journal of Geophysical Research 111, E6, doi:10.1029/2005je002588.

McCleese, D. J., N. G. Heavens, J. T. Schofield, W. A. Abdou, J. L. Bandfield, S. B. Calcutt, P. G. J. Irwin, D. M. Kass, A. Kleinböh, S. R. Lewis, D. A. Paige, P. L. Read, M. I. Richardson, J. H. Shirley, F. W. Taylor, N. Teanby, R. W. Zurek (2010) "Structure and dynamics of the Martian lower and middle atmosphere as observed by the Mars Climate Sounder: Seasonal variations in zonal mean temperature, dust, and water ice aerosols" Journal of Geophysical Research, 115, E12, doi:10.1029/2010je003677.

Michaels, T. I. and S. C. R. Rafkin (2004) "Large-eddy simulation of atmospheric convection on Mars" Quarterly Journal of the Royal Meteorological Society, 130, doi:10.1256/qj.02.169.

Michaels, T. I. (2006) "Numerical modeling of Mars dust devils: Albedo track generation" Geophysical Research Letters 33, doi:10.1029/2006gl026268.

Murphy, J. R., J. B. Pollack, R. M. Haberle, C. B. Leovy, O. B. Toon and J. Schaeffer (1995) "Three-dimensional numerical simulation of Martian global dust storms" Journal of Geophysical Research, 100, E12, 26357-26376, doi:10.1029/95je02984.

Newman, C. E., S. R. Lewis, P. L. Read, and F. Forget (2002) "Modeling the Martian dust cycle 2. Multiannual radiatively active dust transport simulations" Journal of Geophysical Research 107, E12, doi:10.1029/2002je001920.

Pollack J. B. (1982) "Properties of dust in the Martian atmosphere and its effect on temperature structure" Advances in Space Research, 2, 45-56, doi:10.1016/0273-1177(82)90104-1.

Rafkin, S. C. R., R. M. Haberle, and T. I. Michaels (2001) "The Mars Regional Atmospheric Modeling System: model description and selected simulations" Icarus 151, 228-256, doi:10.1006/icar.2001.6605.

Rafkin, S. C. R., R. V. M. Sta. Maria, and T. I. Michaels (2002) "Simulation of the atmospheric thermal circulation of a martian volcano using a mesoscale numerical model" Nature, 419, doi:10.1038/nature01114.

Rafkin, S. C. R. (2009) "A positive radiative-dynamic feedback mechanism for the maintenance and growth of Martian dust storms" Journal of Geophysical Research 114, E01009. doi:10.1029/2008JE003217.

Ye, Z. J., M. Segal, and R. A. Pielke (1990) "A Comparative Study of Daytime Thermally Induced Upslope Flow on Mars and Earth" Journal of the Atmospheric Sciences, 47, 612-628, doi:10.1175/1520-0469(1990)047<0612:ACSODT>2.0.CO:2.

Zurek, R. W. (1982) "Martian great dust storms: an update", Icarus, 50, 288-310, doi:10.1016/0019-1035(82)90127-0.